\documentclass{aa}  

\usepackage{graphicx}
\usepackage{txfonts}
\usepackage{lipsum}
\usepackage{subcaption}         
\usepackage{lscape}             
\usepackage{placeins}           
\usepackage{color}
                                
\begin{document}

   \title{On the Possibility of an Extragalactic Positron Annihilation Signal}


%

   \author{
   Thomas Siegert\inst{1}\fnmsep\thanks{Corresponding author: thomas.siegert@uni-wuerzburg.de}
   \and
   Hiroki Yoneda\inst{2,3,4,5}
   }

   \institute{Julius-Maximilians-Universität Würzburg,
        Fakultät für Physik und Astronomie,
        Institut für Theoretische Physik und Astrophysik,
        Lehrstuhl für Astronomie,
        Emil-Fischer-Str.~31,
        D-97074 Würzburg, Germany
   \and
   The Hakubi Center for Advanced Research, Kyoto University, Yoshida Ushinomiyacho, Sakyo-ku, Kyoto 606-8501, Japan
   \and
   Department of Physics, Kyoto University, Kitashirakawa Oiwake-cho, Sakyo-ku, Kyoto 606-8502, Japan
   \and
   RIKEN Nishina Center, 2-1 Hirosawa, Wako, Saitama 351-0198, Japan
   \and
   Kavli Institute for the Physics and Mathematics of the Universe (WPI), UTIAS, The University of Tokyo, 5-1-5 Kashiwanoha, Kashiwa, Chiba 277-8583, Japan
   }

   \date{Received February XX, 2026}

 
  \abstract
   {With 20 years of INTEGRAL/SPI observations, \citet{Yoneda2025_511} created the most detailed map of the positron annihilation line at 511\,keV. While central bulge and extended disk are readily recognised in this map, several hotspots at high latitude regions may either be imaging artefacts or true signals.}
   {We discuss the possibility of extragalactic positron annihilation signals from hotspots in this map. We also calculate a cosmological positron annihilation signal as a contribution to the Cosmic Gamma-ray Background (CGB).} 
   {For this investigation, we compare 511\,keV emission hotspots away from the Galactic plane with a high velocity cloud column density map as well as with the catalogue of Local Volume Galaxies (LVGs) up to 25\,Mpc.}
   {We find that in particular the Magellanic Stream in the southern and Complex C in the northern sky matches the brightest hotspots, which may indicate a higher positron production rate inside the Milky Way than measured from the Galactic interstellar medium alone, of $10^{44}\,\mathrm{s^{-1}}$. In addition, we can explain other hotspots by the cumulative effect of LVGs in the selected regions. The CGB contribution from positron annihilation might be sub-dominant on the per-cent level. However, depending on the true intrinsic annihilation spectrum, in particular depending on the positron injection energy for in-flight annihilation and the star formation rate per galaxy, a much higher imprint beyond 10\% is possible above several MeV.} 
   {If these findings turn out to be true, next generation MeV telescopes will, for the first time, identify individual extragalactic 511\,keV sources. In particular, several dwarf spheroidal galaxies with fluxes of up to $(1$--$2) \times 10^{-5}\,\mathrm{ph\,cm^{-2}\,s^{-1}}$, the galaxies M31 and M33, as well as some of their satellites, with potentially several $10^{-6}\,\mathrm{ph\,cm^{-2}\,s^{-1}}$, each, may be detected.}


   \maketitle

\section{Introduction}\label{sec:intro}
With more than 20 years of INTEGRAL/SPI \citep{Winkler2003_INTEGRAL,Vedrenne2003_SPI} data, \citet[][hereafter HY25]{Yoneda2025_511} performed a Richardson-Lucy (RL) image reconstruction of the positron annihilation line at 511\,keV.
The strong Galactic bulge emission was seen together with an undoubted signal from the Galactic disk.
In addition to these now well-established emission features, several, also highly significant, emission regions away from the Galactic bulge and plane emerged.
While image reconstruction is often prone to artefacts, systematic tests applied to the dataset, including bootstrapping of the full dataset and background-only image reconstructions, confirm persistency of some emission regions.
These high-latitude ($|b| \gtrsim 15^\circ$) emission features, which may have also been seen with the COSI \citep{Tomsick2024_COSI} balloon experiment in 2016 \citep{Kierans2017_COSI,Siegert2020_511}, demand a thorough discussion and interpretation.

\begin{figure}
    \centering
    \includegraphics[width=1.0\linewidth]{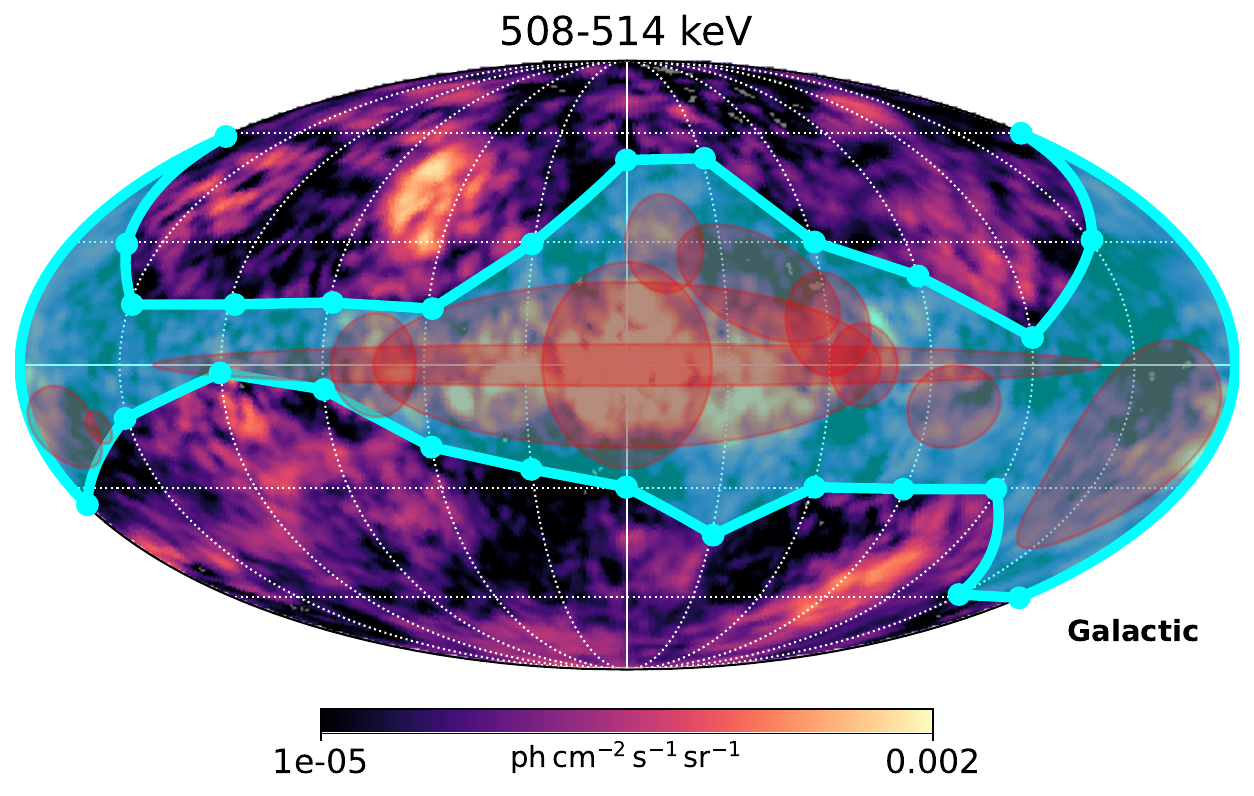}
    \caption{{RL 511\,keV map from HY25 in Mollweide projection with high contrast and reduced lower threshold. The masked region (cyan) is chosen to remove the Galactic bulge, thin and thick disk, nearby massive star regions such as Scorpius Centaurus ($\Delta\ell \approx [290^\circ,360^\circ]$, $\Delta b \approx [-10^\circ,+30^\circ]$), Orion-Eridanus ($[180^\circ,230^\circ]$, $[-50^\circ,-5^\circ]$), Taurus ($[160^\circ,180^\circ]$, $[-25^\circ,-5^\circ]$), Perseus ($[155^\circ,165^\circ]$, $[-25^\circ,-15^\circ]$), Vela ($[245^\circ,280^\circ]$, $[-25^\circ,+5^\circ]$), Carina ($[285^\circ,295^\circ]$, $[-5^\circ,+5^\circ]$), and Cygnus ($[70^\circ,90^\circ]$, $[-10^\circ,+10^\circ]$), plus a margin of a few degrees to take into account the surrounding HI regions. The lower threshold defines the cut regions.}}
    \label{fig:HY_masked_region}
\end{figure}

Nearby extragalactic sources up to a $\sim 15$\,Mpc (redshift $z \lesssim 0.0034${, corresponding to a line shift of $\lesssim 2$\,keV}) may contribute in the imaged 511\,keV bin from 508--514\,keV as the emission line would not be shifted out of this {band}.
Certainly also higher redshifts may contribute, but the line would be shifted towards lower energies and most of the emission would be found outside of the studied bin.
{At distances larger than 15\,Mpc, that is, about the distance to the Virgo Cluster, the cosmological redshift would start moving larger fractions of the $\gamma$-ray line out of the selected window.
For example, at the distance of the Fornax Cluster at $\approx 19$\,Mpc ($z = 0.0046$), the line would already be centred at 508.7\,keV, so at most 50\% of the total line flux would be included from Fornax.
At a distance to the Coma Cluster ($\approx 100$\,Mpc, $z = 0.0231$), the line would be totally shifted out of the 508--514\,keV bin as the centroid would be at 499.5\,keV.}
There are more than 1500 galaxies in the local volume, the so-called Local Volume Galaxies \citep[LVGs;][]{Karachentsev2004_LVG,Karachentsev2005_LVG,Karachentsev2013_LVG}, whose cumulative signal could be within reach of SPI's sensitivity.
Some emission hotspots, even though not statistically significant in themselves, may be coincident with groups of LVGs in the supergalactic plane.
However, the exposure of INTEGRAL is highly non-uniform in the sky, so that not every nearby galaxy group would contribute.
High exposure regions off the Galactic plane are found in the regions of Andromeda, Ursa Major, Hydra/Sextans, and Virgo.
Besides these regions, two more hotspots are found in which hardly any galaxy is found.
Those regions appear to coincide with high velocity clouds \citep[HVCs;][]{Westmeier2018_HVC} and in particular the Magellanic Stream.

In this paper, we assess the reconstructed 511\,keV fluxes around HVCs, the Magellanic Stream, and large accumulations of nearby galaxies (Sec.\,\ref{sec:image_assessment}), and discuss the possibility of them being of astrophysical origin (Secs.\,\ref{sec:mag_stream} \& \ref{sec:exgal}).
Further, we calculate an expected cosmological positron annihilation signal (Sec.\,\ref{sec:cosmo511}) by assuming different dependencies on the star formation rate (SFR), and compare it to current measurements of the Cosmic Gamma-ray Background (CGB).
We summarise our findings in Sec.\,\ref{sec:summary}.

\begin{figure*}[!ht]
    \centering
    \includegraphics[width=0.75\linewidth]{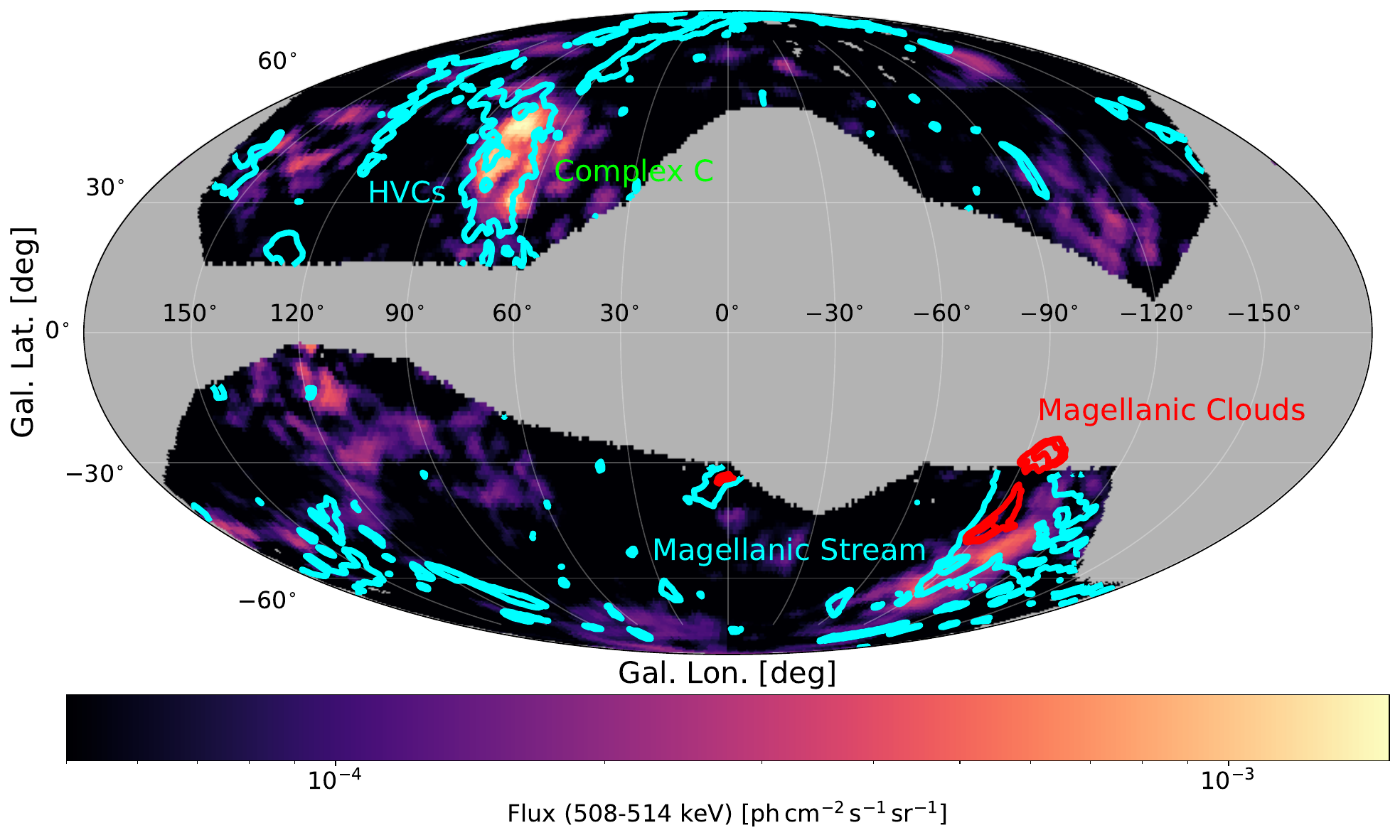}
    \caption{RL 511\,keV map from HY25 with the bright bulge and disk regions masked out (gray) to emphasise on the low surface brightness region at high latitudes. The cyan contours show the HVCs of \citet{Westmeier2018_HVC} at a hydrogen column density of $10^{19}\,\mathrm{cm^{-2}}$. The highest column densities around the Magellanic Clouds ($(\ell,b)_{\rm LMC} = (280.5^\circ,-32.9^\circ)$, $(\ell,b)_{\rm SMC} = (302.8^\circ,-44.3^\circ)$) are indicated in red at level of $10^{20}\,\mathrm{cm^{-2}}$. The two strongest high-latitude 511\,keV regions around $(\ell_0,b_0) \sim (70^\circ, 40^\circ)$ and $(\ell_0,b_0) \sim (-100^\circ, -55^\circ)$ largely coincide with the HVC Complex C and high-density parts of the Magellanic Stream, respectively. Other 511\,keV emission regions may be associated with LVGs (see Fig.\,\ref{fig:regions}).}
    \label{fig:511_vs_MS}
\end{figure*}

\section{Artefact, Galactic or Extragalactic?}\label{sec:image_assessment}
{In Fig.\,\ref{fig:HY_masked_region}, we show the RL 511\,keV map from HY25 in the band 508--514\,keV from 20.5\,yr of INTEGRAL/SPI data with high contrast and reduced lowered flux threshold to emphasise on the high-latitude excesses.
The masked regions include the bulge, the disk, and some nearby emission regions potentially associated with the $\beta^+$-decay of $^{26}\mathrm{Al}$, such as Orion-Eridanus, Perseus, Taurus, Scorpius-Centaurus, Cygnus, Carina, and Vela, as well as large parts of the Galactic anti-centre including the Crab.
In Fig.\,\ref{fig:511_vs_MS}, we then show the same map overlaid with nearby gas clouds.}
HY25 estimated the significance of several individual regions to be between $2$ and $4\sigma$, each, based on a comparison of bootstrapping samples from the entire dataset and background-only datasets.
Several other emission hotspots in this band appear when the bright Milky Way foreground is masked out.
There are at least six regions which appear far from the Milky Way plane.
At least two of them may be associated with the Magellanic Stream and HVCs as indicated in Fig.\,\ref{fig:511_vs_MS}.
The others are coincident with nearby galaxy groups in the regions around Andromeda, Ursa Major, Hydra/Sextans, as well as the Virgo Cluster.

We extract the 511\,keV fluxes from these six regions and combinations thereof by visually setting elliptical boundaries around the hotspots.
While this method may be slightly biased, we will use the bootstrapping samples to estimate confidence intervals for each selected region.
In Sec.\,\ref{sec:exgal}, we take into account only those galaxies which fall exactly into these ellipses so that any in- or decrease of the regions would be properly taken into account.
The details about the regions selected in this study are found in Tab.\,\ref{tab:regions}

\begin{table*}[!ht]
    \centering
    \begin{tabular}{l|rrrrrrrrp{4.75cm}}
        Name & $\ell_0$ & $b_0$ & $a_0$ & $e_0$ & $\theta_0$ & $d_0^{\pm}$ & $d_1^{\pm}$ & $d_2^{\pm}$ & Comments \\
        \hline
        Complex C & $70$ & $40$ & $17.5$ & $0.515$ & $0$ & $0.004$--$0.012$ & -- & -- & HVC \\
        Mag. Stream & $-100$ & $-55$ & $25.0$ & $0.866$ & $-10$ & $0.020$--$0.100$ & -- & -- & -- \\
        \hline
        Andromeda & $121$ & $-21$ & $23.0$ & $0.000$ & $0$ & $0.020$--$0.040$ & $0.52$--$1.31$ & $2.5$--$12.5$ & nearby dSphs, M31 group, NGC\,628/784 extended groups\\
        Ursa Major & $140$ & $40$ & $18.0$ & $0.000$ & $0$ & $0.020$--$0.100$ & $1.37$--$5.55$ & $7.2$--$14.9$ & nearby dSphs, M81 group, NGC\,2787/3556 ext. groups \\
        Hydra & $-115$ & $30$ & $20.0$ & $0.662$ & $25$ & $0.080$--$0.150$ & $1.29$--$1.45$ & $8.2$--$23.0$ & nearby dSphs, NGC\,3109 group, NGC\,3115/2784 ext. groups \\
        Virgo & $-125$ & $70$ & $15.0$ & $0.745$ & $0$ & $0.040$--$0.500$ & -- & $1.7$--$25.2$ & nearby dSphs, galaxies along line of sight, Virgo Cluster \\
        \hline
    \end{tabular}
    \caption{Extragalactic regions with enhanced 511\,keV flux. The columns, from left to right, include an elliptically-shaped region's central Galactic longitude, $\ell_0$, in units of degrees, its Galactic latitude, $b_0$, in degrees, its semi-major axis, $a_0$, in degrees, its eccentricity, $e_0$, its position angle, $\theta_0$, in degrees, three distance range estimates, $d_{0,1,2}^{\pm}$, of known features or sub-groups along the line of sight, in units of Mpc, and comments about associations. The table is split into possible annihilation in gas surrounding the Milky Way (top) and other galaxy groups (bottom).}
    \label{tab:regions}
\end{table*}

\begin{figure}[!ht]
    \centering
    \includegraphics[width=1.0\linewidth]{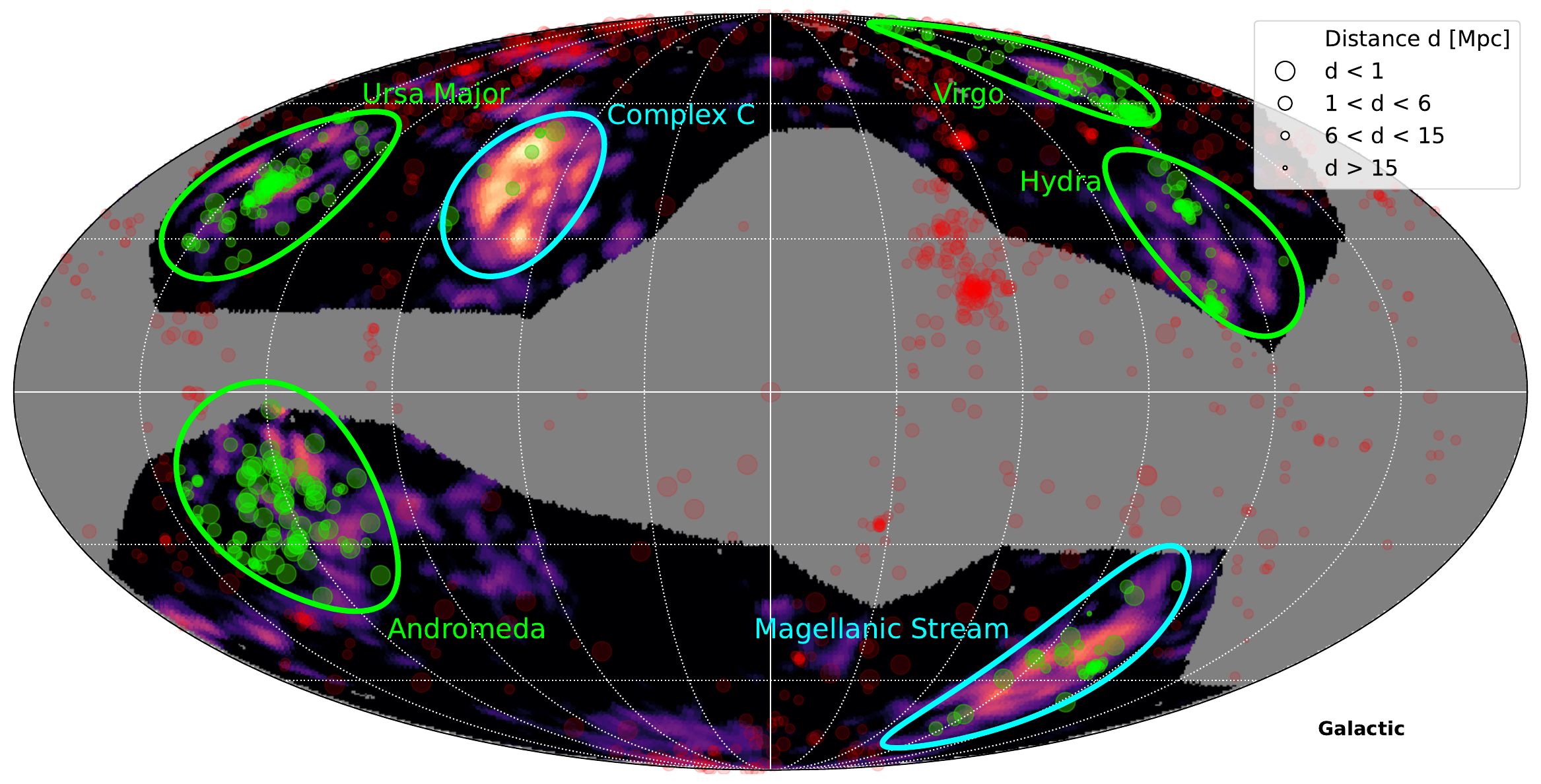}
    \caption{Selected regions in this study (ellipses) on top of the masked HY25 511\,keV map. We show the distribution of all known ($\sim 1600$) LVGs up to a distance of 30\,Mpc in red outside, and green inside the LVG regions (green). Distance intervals are indicated as marker sizes.}
    \label{fig:regions}
\end{figure}

The regions are indicated in Fig.\,\ref{fig:regions} in which the HY25 map is shown again, together with all LVGs up to 30\,Mpc.
Most of these emission regions fall into low-exposure observations or are neighbouring high-exposure regions so that a strong exposure gradient within only a few degrees across the sky is emerging.
While this is naturally handled by proper application of the imaging response function (IRF) in the RL reconstruction algorithm, it might still be the case that the IRF itself is flawed, that the instrumental background modelling approach \citep{Diehl2018_BGRDB,Siegert2019_SPIBG} still produces some systematics, or both.
HY25 tested this masking parts of the IRF which may be affected the strongest, i.e. at large zenith angles beyond the field of view of $16^\circ$, and found that the resulting image is almost identical.
While this is not a direct proof that the response is correct, it shows that at large zenith angles in the far wings of the field of view have no significant impact on the resulting image.
The background model relies on the fact that the geometry of the instrument is not changing with time so that the background ``pattern'' in the energy bin studied only changes whenever a detector fails.
%
%
%
%
Fore more details on background handling, we refer the reader to HY25.

Because the exposure is non-uniform, the sensitivity varies across the map, and not every region can be treated equally.
For this reason, HY25 performed bootstrap sampling by generating many reconstructed images and adding Poisson fluctuations to the dataset, and estimated the uncertainties for a flux from a certain region of interest.
This also allows us to estimate some sort of systematics or signal-to-noise ratio (significance) by taking into account background-only samples.
We note that methods such as likelihood fitting can, in principle, account for correlations between pixels and the background.
If the bootstrap samples were truly independent, this would similarly be ensured.
However, because the present approach adds noise to the real data, the degree to which sample independence is guaranteed is not straightforward.
As a reference, comparing the bootstrap uncertainties with those from a model fitting approach for the total Galaxy flux, we find a factor of 2–-3 difference.
\citet{Siegert2016_511, Siegert2019_511, Siegert2022_511} also showed that neglecting covariance terms can lead to an overestimation of uncertainties by a factor of two or more.
Given these considerations, the bootstrap uncertainties may provide a conservative estimate in this case.
However, since this work focuses on the scientific interpretation and impact in the case of future confirmation rather than a detailed discussion of detection significance, we use the bootstrap uncertainties throughout.

\begin{figure*}
    \centering
    \includegraphics[width=1.0\linewidth]{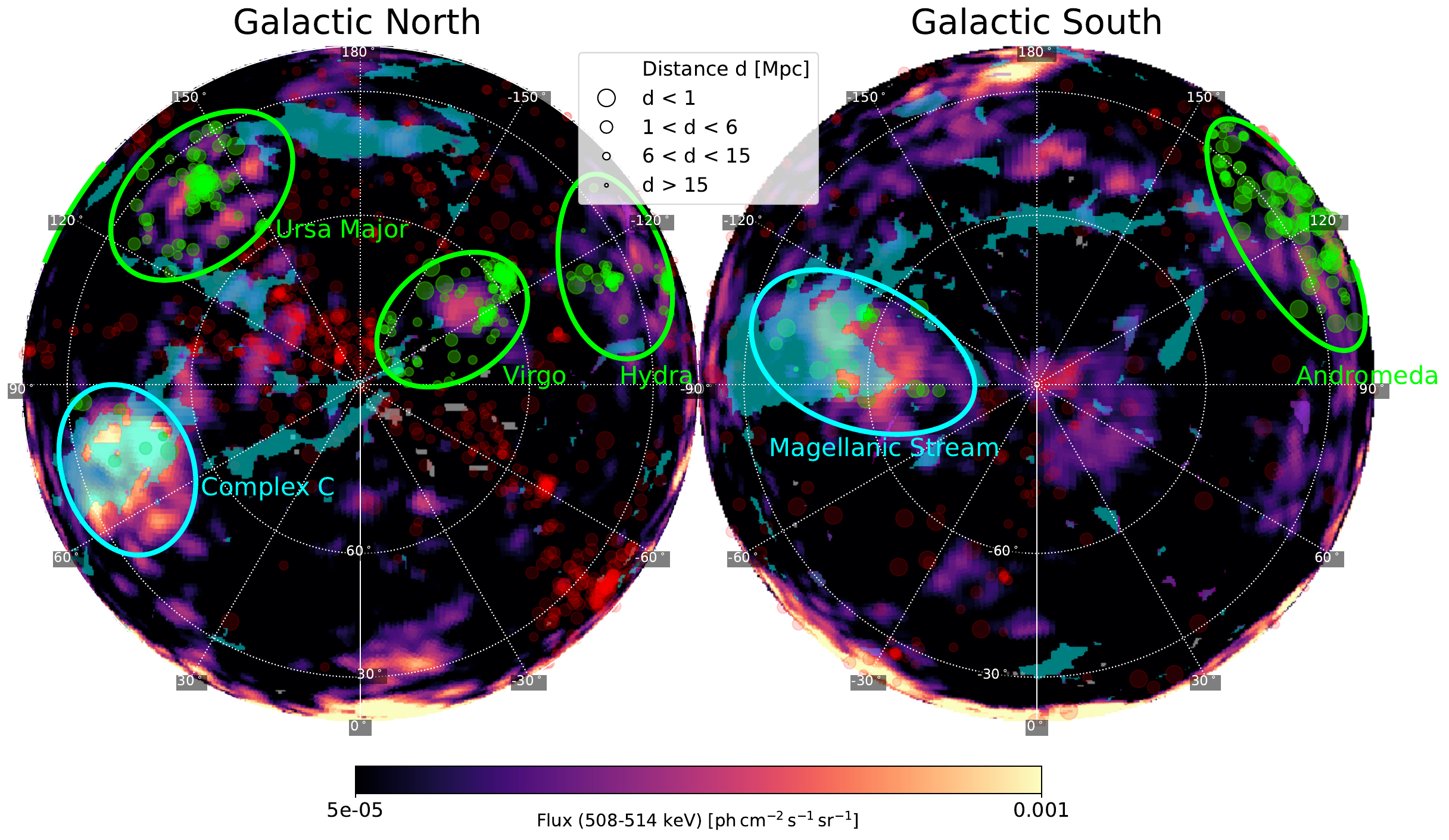}
    \caption{{Same as Figs.\,\ref{fig:511_vs_MS} and \ref{fig:regions}, now combined in orthographic projection to emphasise on the Galactic poles. The Galactic North is shown on the left, the Galactic South on the right. All regions are marked as before, with the Magellanic Stream as cyan overlay. The Galactic centre and bulge are visible as bright region at the bottom of both panels.}}
    \label{fig:all_regions}
\end{figure*}

Given the regions in Tab.\,\ref{tab:regions}, we estimate the 511\,keV fluxes from the HY25 map and their uncertainties in Tab.\,\ref{tab:fluxes}.
\begin{table}[!ht]
    \centering
    \begin{tabular}{c|r|rrrrr}
        Name & $F_{511}$ & $p2.5$ & $p16$ & $p50$ & $p86$ & $p97.5$ \\
        \hline
        Complex C & $9.1$ & $3.6$ & $5.2$ & $7.2$ & $10.3$ & $14.0$  \\
        Mag. Stream & $4.9$ & $2.0$ & $3.0$ & $4.4$ & $6.4$ & $9.0$ \\
        Combined & $14.0$ & $7.2$ & $9.3$ & $12.1$ & $15.5$ & $19.9$ \\
        \hline
        Andromeda & $4.3$ & $3.0$ & $3.9$ & $5.2$ & $7.1$ & $9.5$ \\
        Ursa Major & $2.4$ & $1.2$ & $1.8$ & $2.8$ & $4.4$ & $6.9$ \\
        Hydra & $2.5$ & $1.4$ & $2.0$ & $2.7$ & $3.7$ & $5.4$ \\
        Virgo & $0.8$ & $0.4$ & $0.6$ & $0.9$ & $1.5$ & $2.5$ \\
        Combined & $10.1$ & $8.6$ & $10.1$ & $12.3$ & $14.9$ & $17.8$ \\
        \hline
        All combined & $24.0$ & $18.2$ & $20.9$ & $24.8$ & $28.5$ & $33.5$ \\
        \hline
    \end{tabular}
    \caption{Measured 511\,keV fluxes in the regions of interest as well as combined (summed) fluxes for HVCs (top) and LVGs (bottom). The columns include the integrated fluxes from the HY25 map in the energy bin 508--514\,keV in units of $10^{-5}\,\mathrm{ph\,cm^{-2}\,s^{-1}}$, and the confidence intervals (with $p2.5$ being the 2.5\% percentile, etc.) from the bootstrapping samples in the same units.}
    \label{tab:fluxes}
\end{table}
We show the resulting bootstrap distributions from resampling the dataset and the background-only in Figs.\,\ref{fig:bootstrap} for the combined regions of HVCs and LVGs.
Clearly, the more nearby regions of the HVCs can hardly be explained by background fluctuations.
The extragalactic sources have -- even combined -- a relatively small exposure so that there is no strong evidence for those to not be background fluctuations.
%
%
However, several hotspots appear preferentially in directions towards large galaxy accumulations and nearby gas clouds.
While this is in itself not statistically conclusive, such a spatial coincidence motivates a physical interpretation as discussed in the following {(see Fig.\,\ref{fig:all_regions} for a detailed view towards the Galactic poles)}.
%

\section{Magellanic Stream and High Velocity Clouds}\label{sec:mag_stream}
\citet{Westmeier2018_HVC} used the HI4PI dataset to extract the column densities and velocity profiles for Milky-Way-associated HVCs.
We find that the two major regions in the HVC catalogue, the Magellanic Stream in the southern sky and Complex C in the northern sky, respectively, align well with the strongest 511\,keV emission hotspots at higher latitudes, $|b| \gtrsim 15^\circ$.
Based on the visual inspection in Sec.\,\ref{sec:image_assessment}, we find 511\,keV fluxes of $F_{\rm CC} = (9.1 \pm 2.6) \times 10^{-5}\,\mathrm{ph\,cm^{-2}\,s^{-1}}$ for Complex C and $F_{\rm MS} = (4.9 \pm 1.6) \times 10^{-5}\,\mathrm{ph\,cm^{-2}\,s^{-1}}$ for the Magellanic Stream region, respectively.
The uncertainties are estimated from the bootstrapping samples.
The combined flux of the two regions is $(14.0 \pm 3.1) \times 10^{-5}\,\mathrm{ph\,cm^{-2}\,s^{-1}}$.
From the background-only bootstrap samples (Fig.\,\ref{fig:bootstrap}, left), we estimate the expected background fluctuation to obtain a systematic uncertainty of $4.2 \times 10^{-5}\,\mathrm{ph\,cm^{-2}\,s^{-1}}$.
From the magnitude of those systematics, it appears unlikely that the two emission regions combined stem from background fluctuations.
Given the suggested distances of Complex C of 4--12\,kpc \citep{Wakker2007_ComplexC}, and to the Magellanic Stream of 20--100\,kpc \citep{Mishra2025_MagStream}, the annihilation rates in the two regions are $(0.1$--$4.5) \times 10^{42}\,\mathrm{s^{-1}}$ and $(0.8$--$156.3) \times 10^{42}\,\mathrm{s^{-1}}$, respectively.
This assumes a minimum and maximum distance as well as annihilation through only Ps or only direct.
If the annihilation conditions inside the clouds are similar to the ones in the ISM, the values would be $1$ and $34 \times 10^{42}\,\mathrm{s^{-1}}$, respectively.
The Magellanic Clouds may contribute themselves to the latter value with some unknown fraction.
If the annihilation in the Magellanic Stream is solely fed by the Milky Way, the total Galactic annihilation rate would be at least the canonical value of $5 \times 10^{43}\,\mathrm{s^{-1}}$ \citep{Siegert2016_511} plus the halo / HVC contributions:
%
Considering that those regions only cover about 6\% of the high latitude regions of the sky, a Galactic outflow of positrons may be on the order of $(5$--$7) \times 10^{44}\,\mathrm{s^{-1}}$, i.e. potentially more than ten times stronger than the Galactic value.
Here, we assume isotropic leakage from the Milky Way, which may in reality be affected by the intergalactic magnetic field.
If the annihilation in the Magellanic Stream is fed by processes in the LMC and SMC as major, i.e. most nearby and brightest, contributors in this region and not by the Milky Way alone, the Galactic outflow of positrons would carry on the order of $(3$--$4) \times 10^{43}\,\mathrm{s^{-1}}$.
The latter configuration is more in line with a ``leaky-box model'' of the sub-GeV cosmic-ray content of galaxies with escape fractions of $0.1$--$0.4$ \citep{Lacki2010_CRescape,Yoast-Hull2013_CRescape}, which would be closer to our value of $0.4$--$0.5$ when including only Complex C, rather than $\gtrsim 90\%$ for both regions combined.
However, \citet{Webber2015_CRescape} estimated that, based on the Voyager cosmic-ray electron spectrum, electrons with energies $\lesssim 30$\,MeV may escape in much larger fractions ($\gtrsim 90\%$).
For positrons, this fraction is probably much lower because of the annihilation on the way.
A total Galactic positron production rate based on the measured annihilation rate inside the ISM and in HVCs of $\sim 10^{44}\,\mathrm{s^{-1}}$ seems therefore possible.
We will use this as a canonical value for further discussion.

\section{Cumulative Nearby Extragalactic Signal}\label{sec:exgal}
\begin{figure*}
    \centering
    \includegraphics[width=0.49\linewidth]{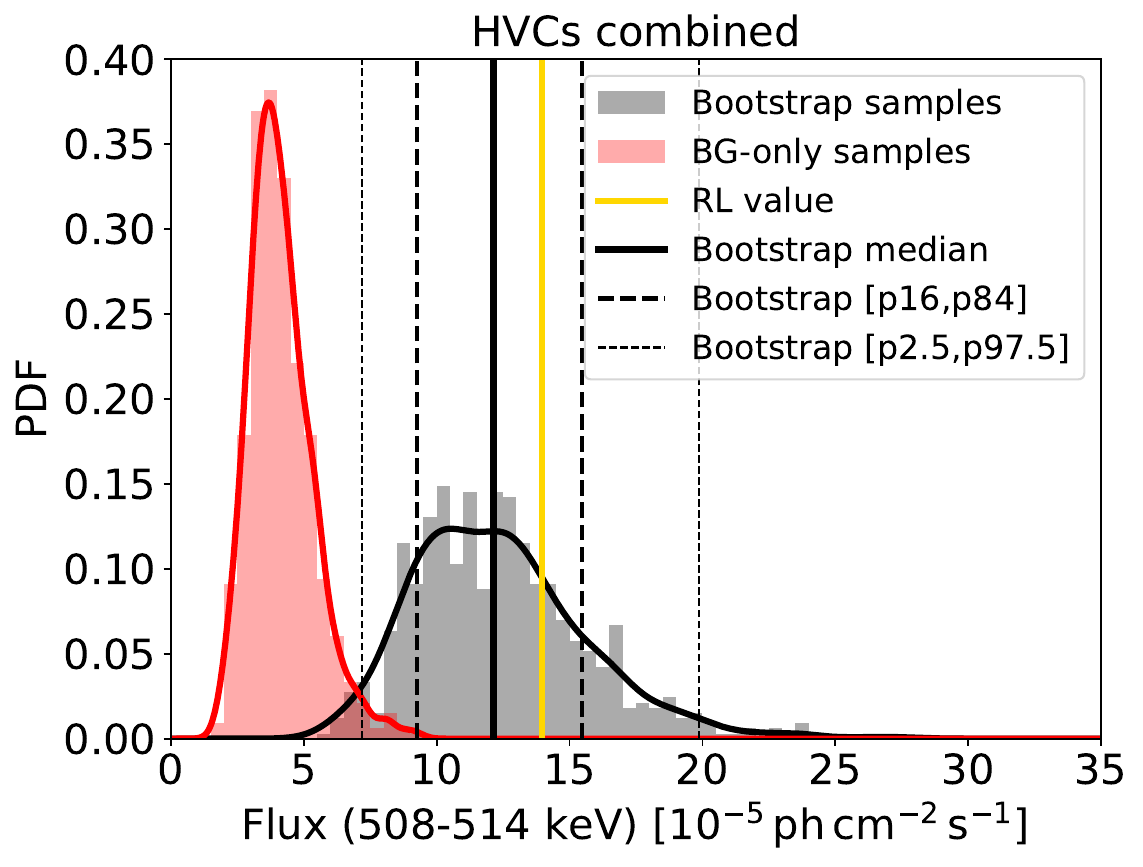}
    \includegraphics[width=0.49\linewidth]{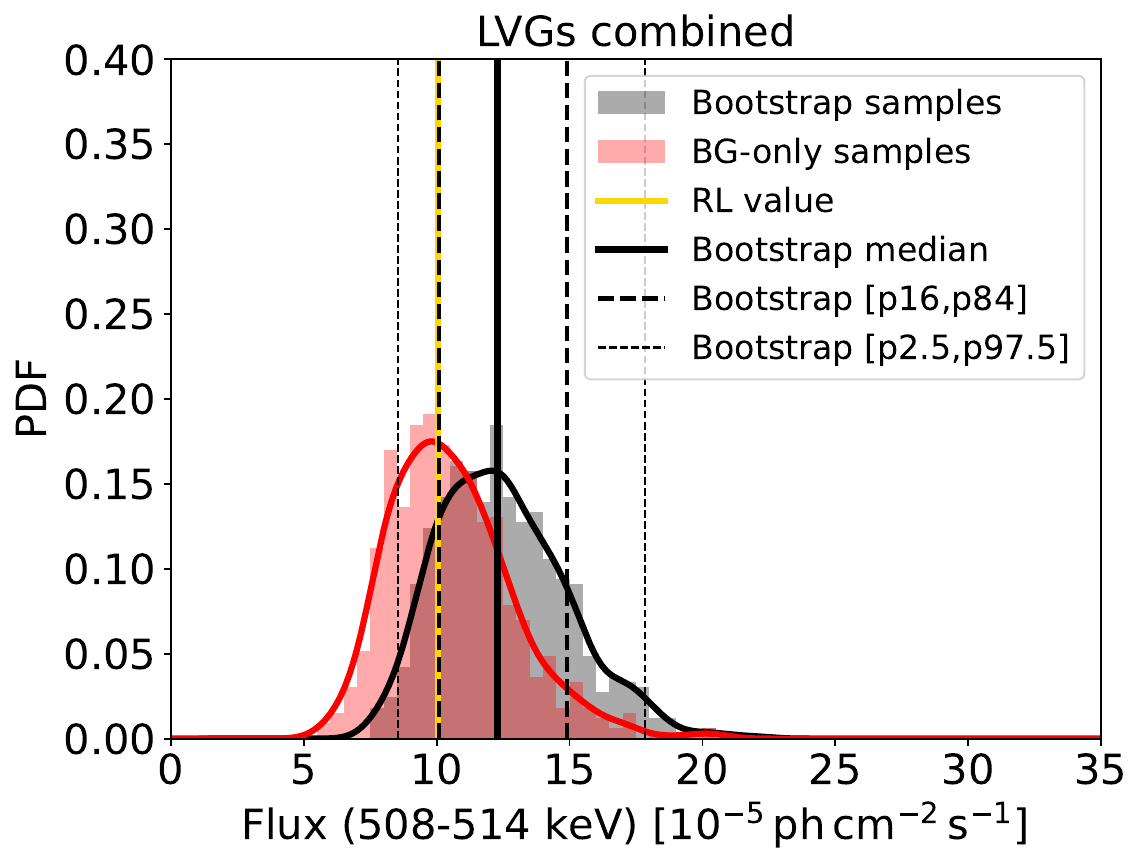}
    \caption{Bootstrapping samples for the combined regions of HVCs (Complex C \& Magellanic Stream; left) and selected LVGs (Andromeda, Ursa Major, Hydra \& Virgo; right). Shown are the samples for the entire dataset (gray histogram) and background-only (red), together with the 68.3 (dashed) and 95.4\% (dotted) confidence intervals and the true RL value (gold vertical line).}
    \label{fig:bootstrap}
\end{figure*}

In addition to the two bright hotspots which may be aligned with HVCs and the Magellanic Stream, there are at least four low surface-bright 511\,keV emission regions.
We find that those align particularly well with nearby galaxy accumulations from the LVG catalogue \citep{Karachentsev2004_LVG,Karachentsev2005_LVG,Karachentsev2013_LVG} of almost 1600 galaxies up to 30\,Mpc.
We name the four regions by their major constellation, Andromeda, Ursa Major, Hydra, and Virgo.
The measured 511\,keV fluxes in the regions are typically very uncertain, with $(4.3 \pm 1.6) \times 10^{-5}\,\mathrm{ph\,cm^{-2}\,s^{-1}}$, $(2.4 \pm 1.3) \times 10^{-5}\,\mathrm{ph\,cm^{-2}\,s^{-1}}$, $(2.5 \pm 0.9) \times 10^{-5}\,\mathrm{ph\,cm^{-2}\,s^{-1}}$, and $(0.8 \pm 0.5) \times 10^{-5}\,\mathrm{ph\,cm^{-2}\,s^{-1}}$, respectively.
In Fig.\,\ref{fig:bootstrap}, right, it becomes evident that the combined flux of those four region in itself is rather significant at $(10.1 \pm 2.4) \times 10^{-5}\,\mathrm{ph\,cm^{-2}\,s^{-1}}$, but that also the systematic uncertainty estimated from the background-only samples is on the order of the flux itself.
There is a high probability that even those four regions combined may be due to background fluctuations.
However, as noted above, the exposure of 20 years of INTEGRAL/SPI observations is quite anisotropic, so that, on one hand, only flux can be detected when the region was actually observed, but on the other hand, the large gradients in the exposure may lead to several artifacts.
We will continue the discussion based on the extracted RL values.

Because the extended extragalactic emission regions may originate as the cumulative effect of many sources at different distances along the line of sight, there is no single annihilation rate per source region.
Instead, the properties of the typically $\sim 100$ LVGs in each region can be used to perform a Bayesian hierarchical decomposition of the measured fluxes into per-source fluxes.
By assuming a simple power-law scaling of the B-band luminosity, $L_{B,i}$, of the sources, and taking into account their distances, $d_i$, we can perform a decomposition by estimating the total source region flux as the sum of the individual LVGs, $F_{i,r}$, as
\begin{equation}
    F_{\mathrm{est},r} = \sum_{i=1}^{N_r} F_{\mathrm{est},i,r} = \sum_{i=1}^{N_r} R_{\rm MW} \frac{\left(L_{B,i}/L_{B,\rm MW}\right)^\gamma}{4 \pi d_i^2}\mathrm{.}
    \label{eq:BHM}
\end{equation}
Here, we normalise by our estimated annihilation rate in the Milky Way, $R_{\rm MW} \sim 10^{44}\,\mathrm{e^+\,s^{-1}}$, and its B-band luminosity, $L_{B,\rm MW} = 2.5 \times 10^{10}\,\mathrm{L}_{B,\odot}$, and fit for the index $\gamma$ to obtain a per-source-flux and therefore annihilation rate.
We assume the observed fluxes, $F_{\rm obs,r}$, per region $r$ to follow a log-normal distribution, $\ln F_{\rm obs,r} \sim \mathcal{N}\left(\ln F_{\mathrm{obs},r},\sigma_{r}^2\right)$, where the total uncertainty per region is $\sigma_{r} = \sigma_{F,r}/F_{\mathrm{obs},r}$.
Each source is assigned some intrinsic scatter, $\sigma_{\rm src}$, so that $\ln F_{i,r} \sim \mathcal{N}(\ln F_{\mathrm{est},i,r}, \sigma_{\rm src}^2)$.
We choose $\sigma_{\rm src} \sim \mathrm{HalfNormal}(0.5)$, which means allowing half an order of magnitude in both direction for the intrinsic scatter of each individual source.
Finally, we set a prior on $\gamma \sim \mathcal{N}(1,1)$ for our parameter of interest.

\begin{figure}
    \centering
    \includegraphics[width=1.0\linewidth]{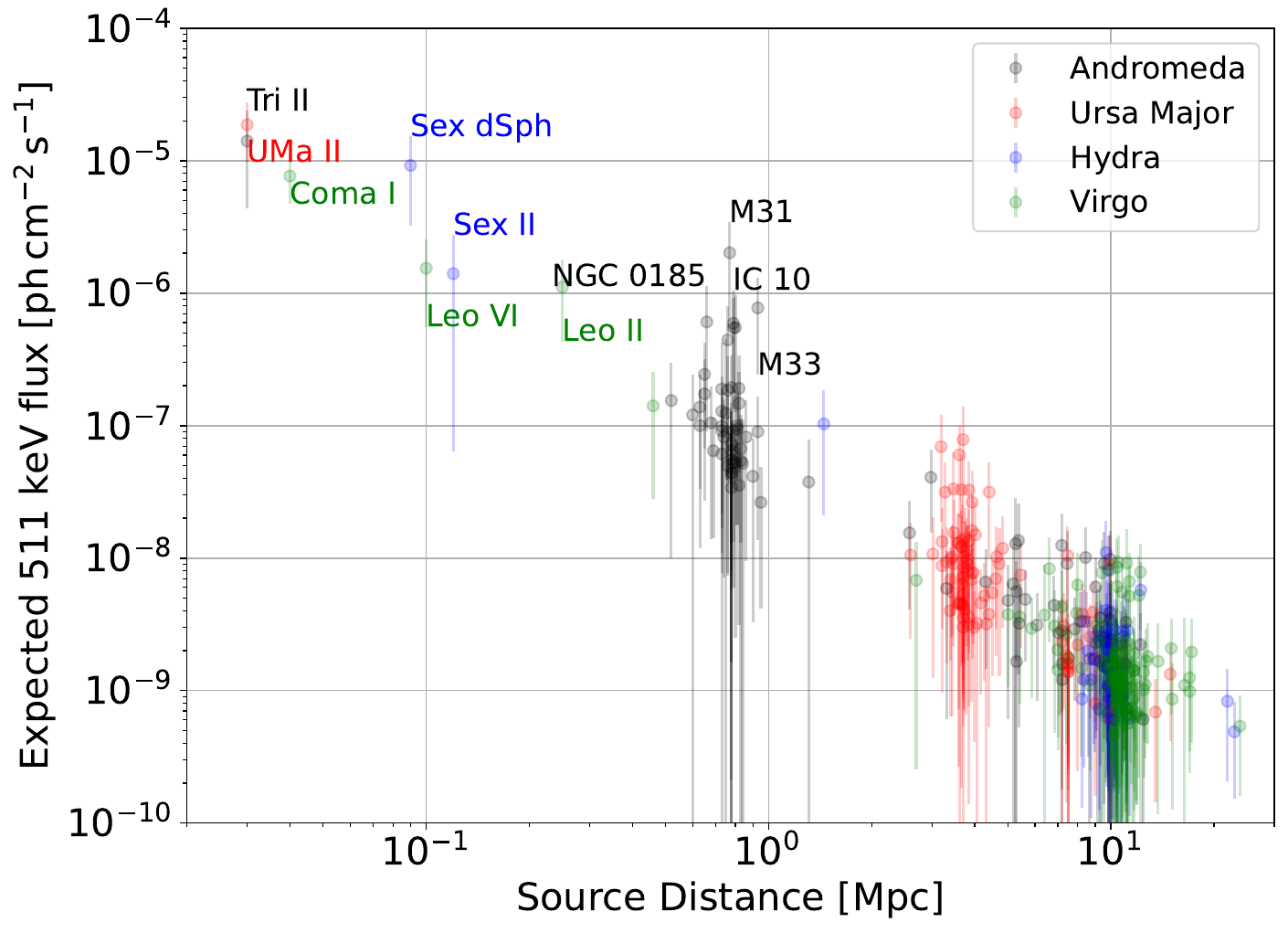}
    \caption{Expected 511\,keV fluxes from individual sources within the bright regions in the study. The fluxes originate from a Bayesian hierarchical decomposition assuming a relation with a galaxy's blue-band luminosity, Eq.\,(\ref{eq:BHM}). Sources with an expected flux stronger than $10^{-6}\,\mathrm{ph\,cm^{-2}\,s^{-1}}$ are named.}
    \label{fig:exgal_fluxes}
\end{figure}

We find a scaling relation like $L_{511} \propto L_B^\gamma$ with $\gamma = 0.25 \pm 0.03$.
%
%
With this scaling, the total predicted fluxes of the four regions are $2.4 \pm 1.0$ for 98 LVG sources, $2.0 \pm 0.9$ for 100, $1.1 \pm 0.6$ for 67, and $(1.1 \pm 0.3) \times 10^{-5}\,\mathrm{ph\,cm^{-2}\,s^{-1}}$ for 154 sources, respectively.
This is somewhat on the low side of the individual regions, indicating potentially missing sources, hidden variables in addition to a scaling with $L_B$, or different annihilation conditions which might lead to larger or smaller line components from positron annihilation.
For each galaxy in our selection, we show the expected 511\,keV flux in Fig.\,\ref{fig:exgal_fluxes}.
All annihilation rates would range between $10^{41}$ and $10^{45}\,\mathrm{s^{-1}}$ for any type of galaxy using this scaling relation.
About ten sources could potentially have a higher flux than $10^{-6}\,\mathrm{ph\,cm^{-2}\,s^{-1}}$, including M31, M33, IC\,10, and NGC\,0185 in the Andromeda region, and four sources more than $10^{-5}\,\mathrm{ph\,cm^{-2}\,s^{-1}}$, all of them being dwarf galaxies with Triangulum\,II and Ursa Major\,II showing the largest fluxes around $2 \times 10^{-5}\,\mathrm{ph\,cm^{-2}\,s^{-1}}$.
Comparing these values with the 511\,keV study from \citet{Siegert2016_dsphs}, all values are still consistent, i.e. below, with previous upper limits.
Clearly, with a more uniform exposure and/or higher sensitivity, those sources may be readily seen with the next generation MeV telescopes, and in particular with COSI \citep{Tomsick2024_COSI}.

\section{Cosmological Positron Annihilation Signal}\label{sec:cosmo511}
If future 511\,keV observations, like those planned with the Compton Spectrometer and Imager \citep[COSI][]{Tomsick2024_COSI}, could confirm the tantalising hints discussed above, such as the total positron production rate in the Milky Way and the 511\,keV emission from LVGs, we can further discuss some first-order expectations of positron annihilation to the CGB:
Beyond 25--200\,Mpc, the cumulative nearby extragalactic signal should morph into a quasi-isotropic, cosmological positron annihilation spectrum.
Then, the redshift $z$ needs to be taken into account and a cosmological model assumed for the positron annihilation background calculation.
By assuming $L_{511} \propto L_B^\gamma \propto \mathrm{SFR}^\gamma$, we can integrate the expected positron annihilation signal in the Universe from galaxies.
While this might not be the only contribution to cosmological positron annihilation as other source classes and intergalactic medium effects could be happening, this calculation serves as a first order of magnitude estimate of positron annihilation to the CGB.

\begin{figure*}[!ht]
    \centering
    \includegraphics[width=1.0\linewidth]{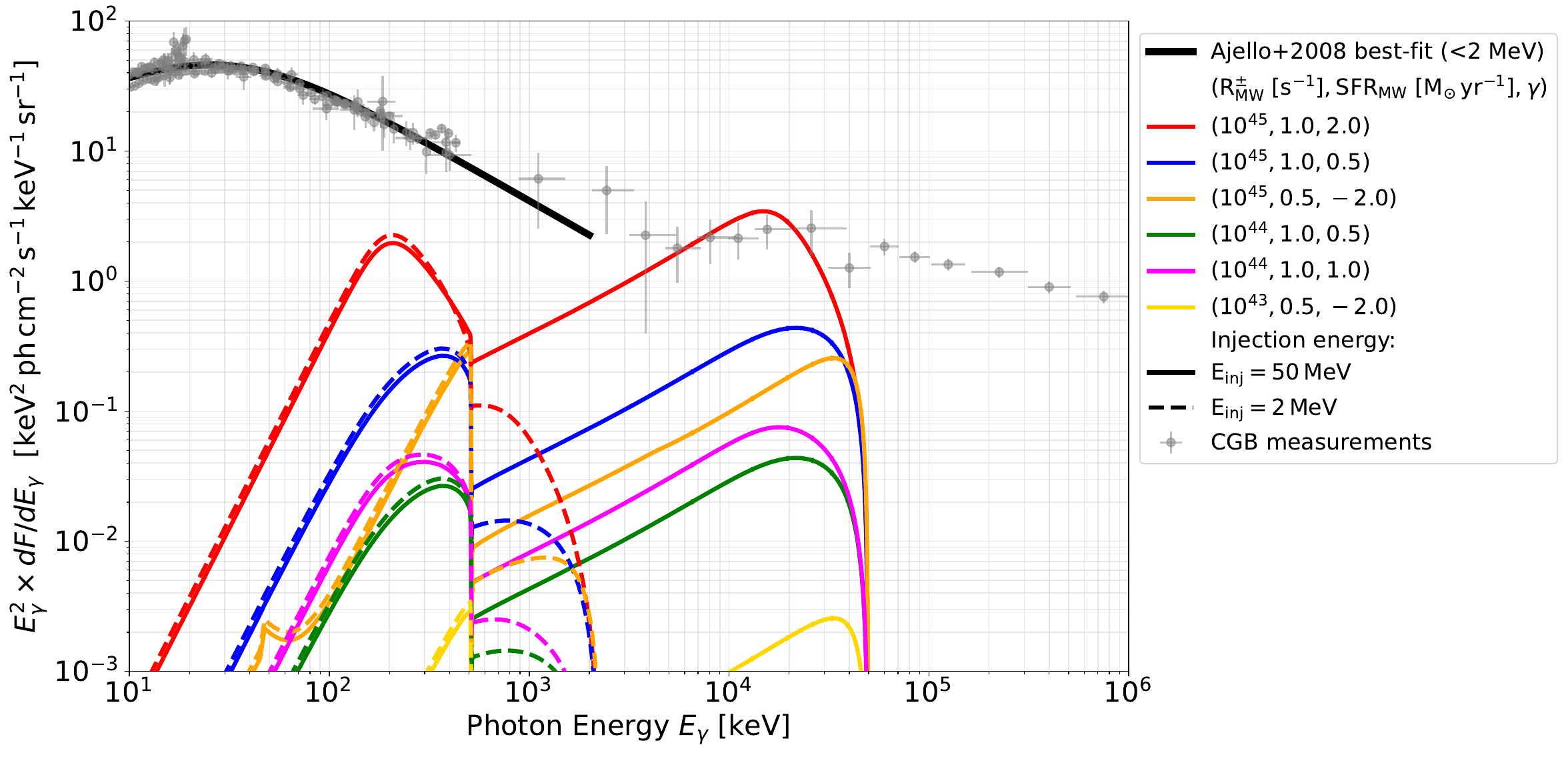}
    \caption{Cosmological positron annihilation signal assuming Eq.\,(\ref{eq:cosmo511}) for different assumptions on the total Galactic positron production (= pair annihilation) rate, $R_{\rm MW}^\pm$, different Galactic SFRs, $\mathrm{SFR}_{\rm MW}$, and dependencies on the SFR history, $\gamma$. Two different but generic positron injection energies are shown. The Ps fraction is set to $f_{\rm Ps} = 0.92$ in both cases. We note that this is not the only contribution to the cosmological positron annihilation spectrum and only assumes some link to SFR. Isotropic Galactic contributions are ignored.}
    \label{fig:cosmo511}
\end{figure*}

The total quasi-isotropic positron annihilation flux from galaxies assuming a relation to their SFR is calculated by
\begin{equation}
    \frac{dI\left(E_{\rm obs}\right)}{dE_{\rm obs}} = \frac{c}{4\pi} \int_{0}^{z_{\rm max}} \frac{\dot{n}(z, R_{\rm MW}, \mathrm{SFR}_{\rm MW}, \gamma) \cdot S_{\rm em}\left(E_{\rm em},f_{\rm Ps},E_{\rm inj}\right)}{H(z) \cdot (1+z)} dz\mathrm{.}
    \label{eq:cosmo511}
\end{equation}
In Eq.\,(\ref{eq:cosmo511}), $\dot{n}(z, R_{\rm MW}, \mathrm{SFR}_{\rm MW}, \gamma)$ is the positron annihilation rate in galaxies as a function of redshift, and defined by scalings of the Milky Way annihilation rate $R_{\rm MW}$ and its SFR $\mathrm{SFR}_{\rm MW}$.
We use
\begin{equation}
    \dot{n}(z, R_{\rm MW}, \mathrm{SFR}_{\rm MW}, \gamma) = \frac{R_{\rm MW}}{\mathrm{SFR_{\rm MW}}} \cdot \rho_{\rm SFR}(z=0)\left(\frac{\rho_{\rm SFR}(z)}{\rho_{\rm SFR}(z=0)}\right)^\gamma
    \label{eq:SFRhistory}
\end{equation}
where
\begin{equation}
    \rho_{\rm SFR}(z) = 0.015 \frac{(1+z)^{2.7}}{1 + \left(\frac{1+z}{2.9}\right)^{5.6}}\,\mathrm{M_\odot\,yr^{-1}\,Mpc^{-3}}
\end{equation}
is the SFR history of the Universe from \citet{Madau2014_SFRH}.
While $\gamma$ is suggested to be close to $0.25$ from Sec.\,\ref{sec:exgal}, the true dependency is unknown and we calculate the expected cumulative signal for a range of $\gamma$ values.
There might also be hidden variables such as the total mass of galaxies, their dark matter content, a delay time distribution to take account the old stellar population, etc., but which might be largely over-interpreting of the current findings.
For this reason, we keep this calculation as simple as possible and refer to future works for more detailed models.
We use a standard cosmological model with a Hubble parameter defined as
\begin{equation}
    H(z) = H_0 \sqrt{\Omega_m(1+z)^3 + \Omega_\Lambda}
\end{equation}
with $\Omega_m = 1-\Omega_\Lambda = 0.315$ and a Hubble constant of $H_0 = 67.4\,\mathrm{km\,s^{-1}\,Mpc^{-1}}$.

We assume a positron annihilation spectrum similar to that of the Milky Way with a 511\,keV line, an ortho-Ps continuum, and annihilation in flight.
For this, we define the following functions:
The survival probability of not annihilating in flight is given by
\begin{equation}
    P_{\rm surv} = P_{\rm surv} \left(E_{\rm inj} \rightarrow \eta \cdot m_e\right) = \exp\left[-n_X J(E_{\rm inj})\right]\mathrm{,}
    \label{eq:P_surv}
\end{equation}
with
\begin{equation}
    J(E) = \int_{\eta \cdot m_e}^E \frac{\sigma(E')}{\left(dE'/dx\right)}\,dE'
    \label{eq:attenuation}
\end{equation}
being the path-integral of the total annihilation cross section along the slowing-down trajectory per unit target density.
In other words, the exponent in Eq.\,(\ref{eq:P_surv}) is an optical depth along the track of positrons.
The factor $\eta$ is set to account for the thermalisation threshold for positrons in the ISM, typically below 1\,keV, so that we set $\eta = 1.001$ in all following calculations.
We note that the choice of $\eta$ does not change the results unless it is set to values much larger than $1.001$.

The number of photons per unit annihilation event is calculated as
\begin{equation}
    r_{\rm Line} = P_{\rm surv} \left( 2 - \frac{3}{2}f_{\rm Ps} \right)
    \label{eq:r_Line}
\end{equation}
for the 511\,keV line,
\begin{equation}
    r_{\rm oPs} = P_{\rm surv} \left(\frac{9}{4}f_{\rm Ps} \right)
    \label{eq:r_oPs}
\end{equation}
for the ortho-Ps continuum, and
\begin{equation}
    r_{\rm IA} = 2 \left( 1 - P_{\rm surv} \right)
    \label{eq:r_IA}
\end{equation}
for the in-flight annihilation component.
In typical ISM conditions where ionisation losses on hydrogen dominate, and for positron injection energies in the MeV range, $P_{\rm surv}$ is between 81\% ($E_{\rm inj} = 50\,\mathrm{MeV}$) and 96\% ($E_{\rm inj} = 2\,\mathrm{MeV}$).
The Ps fraction, $f_{\rm Ps}$, carries the details about the ISM conditions and is typically found to be close to the maximum of $1.0$ in the Milky Way.
We set it to a generic value of $f_{\rm Ps} = 0.92$ given annihilation spectroscopy studies within the last 50 years \citep[see][for a recent review]{Siegert2023_511}.

Assuming natural units ($c=1$), and with a positron mass of $m_e = 511\,\mathrm{keV}$, the generic spectral shapes are for the annihilation line,
\begin{equation}
    L(E_{\rm em}) = \frac{1}{\sqrt{2\pi}\sigma}\exp\left(-\frac{1}{2}\left[\frac{E_{\rm em} - m_e}{\sigma}\right]^2\right)\mathrm{,}
    \label{eq:Line_spec}
\end{equation}
for the ortho-Ps component \citep{Ore1949_oPs}
\begin{eqnarray}
    O(E_{\rm em})
    & = &
    \frac{2}{m_e \left(\pi^2 - 9\right)}\left[\frac{E_{\rm em}(m_e - E_{\rm em})}{(2m_e - E_{\rm em})^2} + \frac{2m_e - E_{\rm em}}{E_{\rm em}} + \right.
    \nonumber\\
    & + & \left.
    \left(\frac{2m_e(m_e - E_{\rm em})}{E_{\rm em}^2} - \frac{2m_e(m_e - E_{\rm em})^2}{(2m_e - E_{\rm em})^3}\right)\right]\mathrm{,}
    \label{eq:oPs_spec}
\end{eqnarray}
and for the in-flight annihilation spectrum
\begin{eqnarray}
A(E_{\rm em})
&=&
\frac{n_e}{1-P_{\rm surv}(E_{\rm inj},\eta \cdot m_e)} \times 
\nonumber\\
& \times &
\int_{E_1(E_{\rm em})}^{E_{\rm inj}}
\frac{P_{\rm surv}(E_{\rm inj},E_+)}{m_e}
\frac{\displaystyle \frac{d\sigma}{dk}\!\left(\gamma(E_+),k(E_{\rm em})\right)}
{\displaystyle \left(\frac{dE_{+}}{dx}\right)\!\left(E_{+},n_X\right)}
\,dE_+\mathrm{,}
\label{eq:IA_spec}
\end{eqnarray}
where we caution that, in general, the electron density, $n_e$, and the density in which the positrons slow down, $n_X$ with $X = \mathrm{H,He,Na,\dots}$ \citep{Panther2018_511}, are \emph{not} necessarily the same.
The reason to use $n_e = n_X$ here, and which was also true in previous studies \citep[e.g.,][]{Churazov2005_511,Beacom2006_AiF,Sizun2006_AiF,Knodlseder2025_AiF}, is that the dependency on the densities cancels in Eq.\,(\ref{eq:IA_spec}), which makes it more analytically tractable.
In fact, for higher positron energies or much lower densities, such as in the intergalactic medium, mostly Inverse Compton and synchrotron losses would shape the in-flight annihilation spectrum, rather than ionisation losses and Coulomb scattering in plasmas.
The differential cross section $\frac{d\sigma}{dk}\left(\gamma(E_+),k(E_{\rm em})\right)$, as well as the integration limits and variable definitions are the same as in \citet{Beacom2006_AiF}.

The spectral functions are normalised to $1.0$,
\begin{equation}
    \int L(E)\,dE = \int O(E)\,dE = \int A(E)\,dE = 1\mathrm{.}
    \label{eq:spec_normed}
\end{equation}
The total emission spectrum is then
\begin{equation}
    S_{\rm em}\left(E_{\rm em},f_{\rm Ps},E_{\rm inj}\right) = r_{\rm Line}L(E_{\rm em}) + r_{\rm oPs}O(E_{\rm em}) + r_{\rm IA}A(E_{\rm em})\mathrm{,}
    \label{eq:spec_function}
\end{equation}
so that integrating over the photon energy gives
\begin{equation}
    \int S_{\rm em}\left(E_{\rm em}\right)\,dE = r_{\rm Line} + r_{\rm oPs} + r_{\rm IA} \equiv N_{\gamma}\mathrm{,}
    \label{eq:number_of_photons}
\end{equation}
which is simply the average number of photons emitted per electron-positron annihilation event.
This leads to spectra as shown in Fig.\,\ref{fig:cosmo511}.

Given that the true dependency on the SFR, and that also other parameters, are unknown, one could constrain those by measuring the CGB to a much higher accuracy as currently achieved.
We note that there would also be an isotropic 511\,keV line as well as the other annihilation components from the Milky Way itself as the observer sits inside.
This would add to cosmological isotropic component, but is also beyond the scope of this work.

If the total positron annihilation rate of the Milky Way is around $10^{44}\,\mathrm{s^{-1}}$ or above, which may be suggested by this work, some imprint in the CGB could be expected already from this component.
With the future launch of COSI in 2027, the CGB can be measured with a much higher accuracy in the MeV range so that any hints of a sharp cut-off at 511\,keV, a rising spectrum below 511\,keV, or a rising spectrum in the 1--5\,MeV range may point to a cosmological positron annihilation component.
Clearly, other long-sought CGB components, such as the cumulative cosmological spectrum of type Ia supernovae \citep[e.g.,][]{Ruiz-Lapuente2016_CGB}, kilonovae \citep[e.g.,][]{Ruiz-Lapuente2020_CGB}, IGM processes (Gelowicz et al. 2026, in prep.), and the Milky Way \citep{Siegert2023_511} and Local Bubble \citep{Siegert2024_LocalBubble} foreground will add so that only with detailed modelling and proper statistical data analysis, deciphering the different components will be possible.

\section{Summary and Conclusions}\label{sec:summary}
We analysed the high-latitude regions of the HY25 511\,keV map and found six emission hotspots with a combined flux of $(24.0 \pm 3.8_{\rm stat} \pm 14.3_{\rm syst}) \times 10^{-5}\,\mathrm{ph\,cm^{-2}\,s^{-1}}$.
We separated this flux into regions coincident with HVCs and the Magellanic Stream showing a flux of $(14.0 \pm 3.1_{\rm stat} \pm 4.2_{\rm syst}) \times 10^{-5}\,\mathrm{ph\,cm^{-2}\,s^{-1}}$, and coincident with LVGs with a flux of $(10.1 \pm 2.4_{\rm stat} \pm 10.1_{\rm syst}) \times 10^{-5}\,\mathrm{ph\,cm^{-2}\,s^{-1}}$.
The statistical and systematic uncertainties are estimated from bootstrap sampling of the entire dataset and background-only realisations.
Based on these numbers and the coincident regions, we suggest that the Milky Way might show significant outflows of material carrying positrons, which would enhance the total positron production rate up to $10^{44}\,\mathrm{s^{-1}}$, which is about a factor of 2--3 higher than from measurements inside the interstellar medium.
A possible cumulative extragalactic positron annihilation signal from nearby galaxy groups, such as around Andromeda, Ursa Major, Hydra/Sextans, and Virgo might explain the remaining signal.
We perform a Bayesian flux decomposition of the emission regions using all known LVGs up to a distance of 30\,Mpc and find consistency if the galaxies' 511\,keV flux scales with their B-band luminosity as $\propto L_{B}^{0.25\pm0.03}$.
Then, all galaxies would show intrinsic annihilation rates between $10^{41}$ and $10^{45}\,\mathrm{s^{-1}}$.
Based on this decomposition, several sources with 511\,keV fluxes above $10^{-6}\,\mathrm{ph\,cm^{-2}\,s^{-1}}$, including Triangulum\,II, Ursa Major\,II, Sextans dwarf spheroidal, M31, and M33, among others, might be identified as individual sources with next generation telescopes, such as COSI \citep{Tomsick2024_COSI}.
Taking this model one step ahead, we calculate the cosmological contribution of positron annihilation in galaxies, assuming some dependency on the SFR.
We estimate that this contribution is probably sub-dominant on the per-cent level, however might show some imprints on the ten per-cent level if the intrinsic positron injection energy is tens of MeV or higher.
We note that this is probably not the only cosmological or isotropic positron annihilation component, and we will address this in forthcoming works (Gelowicz et al. 2026, in prep.).
While there is still a chance that \emph{some} high-latitude 511\,keV emission features are imaging artefacts, it appears at odds if \emph{all} hotspots are mere background fluctuations.
%


\bibliographystyle{aa}
\bibliography{thomas}






\end{document}